\documentclass[11pt]{article}
\usepackage{hyperref}
\pdfoutput=1
\begin{document}
\title{Spontaneous vortex generation at 3-phase \\contact line}
\author{Tae-Hong Kim and Ho-Young Kim \\
\\\vspace{6pt} School of Mechanical and Aerospace
Engineering, \\ Seoul National University, Seoul 151-744, Republic of Korea}
\maketitle
\begin{abstract}
This article describes the fluid dynamics video for "Spontaneous vortex generation at 3-phase contact line" presented at the 64th Annual Meeting of the APS Division of Fluid Dynamics in Baltimore, Maryland, November 20-22, 2011.
\end{abstract}
\section{Introduction}
Drop deposition on a substrate gives rise to the motion of a contact line where the three phases of drop/substrate/gas meet. Although the motion of the three-phase contact line on a solid substrate has been extensively studied thus far, the understanding of the dynamics of the contact line of liquid/liquid/gas phases is far from complete. Here we deposit a drop of isopropyl alcohol (IPA) on water whose free surface is exposed to air to observe the flow field around the contact line. By combining the shadowgraph and high-speed imaging techniques, we find that vortices are spontaneously generated at the contact line, which grow in size with time. The flow is attributed to the Marangoni stress that pulls a liquid of lower-surface tension (IPA) toward water surface having a higher surface tension. However, it is not still clear why the entrained lower-surface-tension liquid should whirl rapidly beneath the contact line. We also visualize the flow by the particle image velocimetry (PIV) to find out that the rotational velocity reaches the maximum of 5 cm/s near the free surface.
\end{document}